\documentclass[aps,preprint]{revtex4}

\usepackage{amsmath}\usepackage{amssymb}
\usepackage{epsfig}
\usepackage{graphicx}
\usepackage{bm}

\newcommand{\be}{\begin{equation}}
\newcommand{\dd}{\displaystyle}
\newcommand{\ee}{\end{equation}}
\newcommand{\bea}{\begin{eqnarray}}
\newcommand{\eea}{\end{eqnarray}}

\begin{document}

%\hfill{\bf BARI-TH 461/03~~~~\,}\vskip0.1cm\hfill{\bf
%DFF-412/03/04\,\,}
\preprint{{\bf BARI-TH 519/05}}

\author{M. Ruggieri}
\affiliation{Universit\`a di Bari, I-70126 Bari, Italia \\and
\\I.N.F.N., Sezione di Bari, I-70126 Bari, Italia}
\title{Smeared gap equations in crystalline color superconductivity}


\begin{abstract}
In the framework of HDET, we discuss an averaging procedure of the
NJL quark-quark interaction lagrangian, treated in the mean field
approximation, for the two flavor LOFF phase of QCD. This procedure
gives results which are valid in domains where Ginzburg-Landau
results may be questionable. We compute and compare the free energy
for different LOFF crystalline structures.
\end{abstract}

\maketitle

%%%%%%%%%%%%%%%%%%%%%%%%%%%%%%%%%%%%%%%%%%%%
%% MAINMATTER
%%%%%%%%%%%%%%%%%%%%%%%%%%%%%%%%%%%%%%%%%%%%

\section{Introduction and the LOFF state}
The behaviour of QCD at very high baryon density and low temperature
has recently attracted a lot of interest. In these conditions,
quarks are expected to deconfine~\cite{Collins:1974ky} and occupy
(in momentum space) large Fermi spheres. At very high densities the
relevant interaction is the one gluon exchange, which is attractive
in the antisymmetric $\bar{\bf 3}$ color channel. Thus, a
Cooper-like pairing phenomenon is expected to occur. This is color
superconductivity~\cite{Barrois:1977xd,Bailin:1983bm,Alford:1997zt,Rapp:1997zu,Schafer:1999jg,Alford:1998mk,
Shovkovy:2003uu,Alford:2003fq} (see
\cite{Rajagopal:2000wf,Alford:2001dt,Nardulli:2002ma,Schafer:2003vz,Schaefer:2005ff}
for reviews).

Regimes of high  baryon densities and low temperatures are expected
to be realized in the core of compact stars; therefore these
superdense objects could be the places where color superconductivity
is realized in nature. Apart the astrophysical speculations, the
study of color-superconductive QCD is an intriguing challenge itself
because its understanding implies a deeper understanding of the QCD
phase diagram.

In nature, as a consequence of electrical and color neutrality and
of the different masses, the Fermi momenta of the quarks should
depend on their color and their flavor. When the difference of the
Fermi momenta of the pairing quarks are too different, a
superconductive ground state in which the Cooper pairs have a net
momentum is energetically favored with respect to the usual, zero
momentum state. The resulting state is known as LOFF phase, and has
been studied in the sixties in the context of condensed matter
physics~\cite{LOFF1,LOFF2}. Possible realizations of the LOFF state
in the frame of color superconductivity have been considered for the
first time in Refs.~\cite{Alford:2000ze,Bowers:2001ip,Bowers:2002xr}
(see~\cite{Casalbuoni:2003wh} for a review and
Refs.\cite{Giannakis:2004pf,Giannakis:2005sa,Giannakis:2005vw} for
recent developments).

A color-superconductive phase is characterized by a nonzero
expectation value of a bilinear quark operator, namely a condensate.
In the LOFF phase with two flavors ($u$ and $d$) one has
\begin{equation}
\langle\psi_{\alpha i}\,C\,\gamma_5\,\psi_{\beta j}\rangle \propto
\Delta({\bm r})\,\epsilon^{\alpha\beta 3}\,\epsilon_{ij 3}~,
\label{eq:ansaztLOFF}
\end{equation}
where $\alpha,\beta$  are color indices and $i,j$ denote the flavor.
A three flavor case ($u$, $d$ and $s$) has been recently considered
in Ref.~\cite{Casalbuoni:2005zp}. As for the space dependence of the
condensate one usually decomposes $\Delta({\bm r})$ as a sum of
plane waves. The most simple ansatz is~\cite{LOFF2}: \be \Delta(\bm
r)=\Delta~ e^{2i{\bf q}\cdot{\bf r}}~, \label{eq:FFansatz}\ee and
the corresponding superconductive phase is known as FF state. The
net momentum of the Cooper pair is $2{\bf q}$, which is the same for
all the pairs while its direction is chosen spontaneously.

One of the most important problem is to calculate the gap parameter
$\Delta$ in Eq.~\eqref{eq:FFansatz}. For the FF state this problem
can be solved exactly: one has simply to shift the momenta of the
paired quarks, ${\bf p}_u\rightarrow{\bf p}_u + {\bf q}$, ${\bf
p}_d\rightarrow{\bf p}_d - {\bf q}$ so the ${\bf q}$ dependence of
the gap parameter disappears. Then one can write a self-consistency
and exactly soluble equation for $\Delta$. In the context of QCD
this procedure has been applied in Ref.~\cite{Bowers:2001ip} where a
diagrammatic approach to the FF color superconductive state is
considered. Unfortunately, this procedure cannot be applied to the
general case of the linear combination of $N$ plane waves. To
overcome this difficulty, one can employ a Ginzburg-Landau (GL)
expansion of the general LOFF free energy functional $\Omega$. GL
expansion works where $\Delta$ is small when compared to the typical
mass scales of the model, in this case $q$ and $\delta\mu$. In the
framework of the GL expansion, in Ref.~\cite{Bowers:2002xr} it is
conjectured that a face centered cube is a good candidate for the
crystalline color superconductive state.

In this paper we wish to define a procedure that allows to study the
crystalline color superconductor with a generic crystalline
structure in a space of parameters which is complementary to the GL
one. This procedure is based on a weighted average of the
interaction lagrangian over the lattice cell. The paper is organized
as follows: in Sec.~\ref{sec:EG} we discuss the smearing procedure
used to obtain the effective gap equations, in the simple case of
the FF state.  Sec.~\ref{sec:GCS} is devoted to the generalization
of this procedure to a generic crystalline structure. In
Sec.~\ref{sec:NR} we show one result of the solution of the smeared
gap equations, namely the free energy plots for the different
crystalline structures. Finally, in Sec.~\ref{sec:CO} we summarize
the leading results of our work.

\section{Smeared lagrangian in the FF phase}\label{sec:EG}
We shall consider ~Cooper pairing of the massless quarks up $u$ and
down $d$, with chemical potential $\mu_u$, $\mu_d$. We define
$\mu=(\mu_u+\mu_d)/2$ and $\delta\mu=|\mu_u-\mu_d|/2\ll\mu$. We
begin with a review of the FF state. Although this case can be
solved exactly, it is useful to consider it here in order to fix the
notations and introduce some definitions to be used later on.

We work in the framework of High Density Effective Theory (HDET)
\cite{Nardulli:2002ma,Hong:1998tn,Hong:1999ru,Beane:2000ms,Schafer:2003jn,Casalbuoni:2001dw,Schafer:2003yh}.
The lagrangian for free quarks can be written as  \be {\cal L}_{0} =
\sum_{\vec v} \Big[\psi_+^\dagger iV\cdot
\partial\psi_+\ +\
\psi_-^\dagger i\tilde V \cdot \partial\psi_-\Big]\ +\ (L\to
R)\,.\label{eq:L0}\ee Here the sum represents an average over
velocities; $\psi_\pm\equiv \psi_{\bf\pm v}$ are velocity dependent,
positive energy left handed fields (the negative energy part has
been integrated out). $\psi_{\bf v}$ depends on the residual
momentum $\ell$, corresponding to the decomposition of the quark
momentum
 $p=\mu v+\ell$, with
$v^\mu=(0,{\bf v})$ and $\ell_\parallel={\bm \ell\cdot}{\bf v}=\xi$.
We also introduce $V^\mu=(1,{\bf v})$ and $\tilde V^\mu=(1,\,-{\bf
v})$.

Next we turn to the interaction term. We consider a Nambu-Jona
Lasinio (NJL) inspired four fermion interaction to mimic the one
gluon exchange of QCD, namely~\cite{Alford:2000ze,Casalbuoni:2004wm}
\be {\cal L}_I=-\frac 3 8 G\bar\psi\gamma^\mu
\lambda_a\psi\,\bar\psi\gamma^\mu \lambda_a\psi\label{eq:291}\,.\ee
Here $G$ is a coupling constant, with dimension {\em mass}$^{-2}$;
$\lambda_a$ are color matrices and a sum over flavors is understood
(the FF state with quark-quark interaction mediated by one gluon
exchange has been considered in Ref.\cite{Leibovich:2001xr}). In the
mean field approximation, after Fierzing, we get \be{\cal
L}_{I}=-\frac 1 2\epsilon_{\alpha\beta
3}\epsilon^{ij}(\psi_i^\alpha\psi_j^\beta\,\Delta(\bm r)+\,{\rm
c.c.})\,+\,(L\to R)-\frac 1 G\Delta(\bm r)\Delta^*(\bm
r)\,,\label{8}\ee where $i,j$ are flavor indices and $\alpha,\beta$
are color indices. In the FF state the total momentum of the Cooper
pair is $2\bf q$ and the condensate has the space-dependence of a
single plane wave, see Eq.~\eqref{eq:FFansatz}. In the HDET
formalism Eq.~\eqref{8} can be recast in the form
 \bea
{\cal L }_{I}&=&-\frac{\Delta} 2\,\sum_{\bf v_i,\bf v_j} \exp\{i{\bf
r}\cdot{\bm\alpha}(\bf v_i,\,\bf v_j,\,{\bf q}
)\}\epsilon_{ij}\epsilon_{\alpha\beta 3}\psi^T_{\bf
v_i;\,i\alpha}(x)C \psi_{-\,\bf v_j;\,j\beta}(x)\cr && -(L\to
R)+{\rm h.c.}-\frac 1 g\Delta(\bm r)\Delta^*(\bm r)\
,\label{loff60}\eea where \be{\bm\alpha}({\bf v_i},\,{\bf
v_j},\,{\bf q})=2{\bf q}-\mu_i{\bf v_i}-\mu_j{\bf v_j}\,.
\label{Lcond}\ee Eqs.~\eqref{eq:L0} and~\eqref{Lcond} are the HDET
lagrangian of paired quarks in the FF state. Armed with them one can
write a gap equation, namely a self-consistence (Schwinger-Dyson)
equation for the gap parameter $\Delta$.

Now we have the ingredients necessary to define our smearing
procedure. To this end we recall the exact FF gap
equation~\cite{Casalbuoni:2004wm}, \be
\Delta=i\,\frac{g\rho}{2}\int\frac{d{\bf v}}{4\pi} \int_0^{\delta}
\frac{d \xi}{2\pi} \int d \ell_0  \,\,
\frac{\Delta_{eff}}{\ell_0^2-\xi^2-\Delta^2_{eff}}
=\frac{g\rho}{2}\int
 \frac{d\,{\bf v}}{4\pi}
 \int_0^\delta\,d\xi\,\frac{\Delta_{eff}}{\sqrt{\xi^2+\Delta^2_{eff}}}
\label{GAPeff} \, , \ee where we have defined an effective gap
parameter, \bea
\Delta_{eff}\,=\Delta\theta(E_u)\theta(E_d)\,=\,{\dd\Bigg \{ }
\begin{array}{cc}\Delta & \textrm{~~for~~}
(\xi,{\bf v}) \in PR\cr&\cr 0 &
\textrm{~~elsewhere~~} \ ,\\
\end{array}
\label{GAP8.1} \eea and $E_{u,d}$ are the dispersion laws for $u$
and $d$ quarks respectively~\cite{Alford:2000ze,Casalbuoni:2003sa}
\be E_{d,u}=\pm\delta\mu\mp {\bf q} \cdot {\bf
v}+\sqrt{\xi^2+\Delta^2}~; \label{dispersionFF}\ee $PR$ denotes the
pairing region, \be\label{PRFF} PR = \left\{ (\xi,{\bf v}) \,|\,
E_u>0 \,\mathrm{and}\, E_d>0 \right\} \, .\ee We stress that
Eq.~\eqref{GAPeff} is exact. The key observation is that we can
obtain the {\em same} gap equation~\eqref{GAPeff} in the framework
of HDET by defining a weighted smearing procedure of the gap
lagrangian~\eqref{loff60} over the lattice cell. First of all, we
note that in the gap equation the relevant momenta are small with
respect to the gap which is of the order of $q$. Therefore we may
assume that the velocity dependent fields are slowly varying over
regions of the order of the lattice size. This means that in the
average we can treat them as constant, and in conclusion the average
is made only on the coefficient $ \exp\{i{\bf r}\cdot{\bm\alpha}\}$.
Therefore what we are computing is \be I ({\bm\alpha})=\Big<
\exp\{i{\bf r}\cdot{\bm\alpha}\}g_R({\bf r})\Big>\ \label{II}\ee
where the bracket means average over the cell, and the weight
function $g_R({\bf r})$ can be chosen in such a way that \be I
({\bm\alpha})=
\delta_R^3\left(\frac{\bm\alpha}{2q}\right)~,~~~~~{\text{where}}
~~~~~\delta_R^3({\bf x})= {\dd\Bigg \{ }
\begin{array}{cc} {1} & \textrm{~~for~~}
|{\bf x}| < \dd{\frac{\pi}{2R}} \, , \cr&\cr
0 & \textrm{elsewhere}\,  \\
\end{array}
\label{deltaII}\ee and $R/\pi \approx 1$. Independently of the exact
form of $ g_R({\bf r})$, we will assume that the average procedure
gives as a result the brick-shaped function $\delta_R$ defined in
(\ref{deltaII}). As shown in Ref.~\cite{Casalbuoni:2004wm}, choosing
\be R=\frac{\pi|\delta\mu-{\bf v\cdot\bf
q}|}{2\sqrt{\xi^2+\Delta^2}|h({\bf v\cdot\hat q})|} ~,~~~~~h({\bf
v\cdot\hat q})=1-\frac{z_q}{\bf v\cdot\hat q}\label{eq:31} \, , \ee
one obtains from Eq.~\eqref{loff60} the smeared lagrangian \be {\cal
L}_{I}=-\frac{1} 2\, \,\sum_{\vec
v}\Delta_{eff}\,\epsilon_{ij}\epsilon_{\alpha\beta 3}\psi^T_{{\bf
v};\,i\alpha}(\ell)C \psi_{-\,{\bf v};\,j\beta}(-\ell)-(L\to R)+{\rm
h.c.}-\frac 1 g\Delta\Delta^* \, ,\label{loff60bis}\ee from which
the desired gap equation~\eqref{GAPeff} is obtained (the fermion
propagator can be read in Eqs.~(22) and (23) of
Ref.~\cite{Casalbuoni:2004wm}). Since $R/\pi\approx 1$, then
$\Delta$ has not to be small, meaning that we should be far from a
second order phase transition.

Let us finally notice that  in Eq.~\eqref{GAPeff} one can perform
the $\ell_0$ integration  by substituting the previous expression
for $R$ with \be R=\frac{\pi|\delta\mu-{\bf v\cdot\bf
q}|}{2|\ell_0|\cdot|h({\bf v\cdot\hat q})|}\ .\label{R}\ee In fact,
observing that in any case $\Delta_{eff}$ is equal to 0 or $\Delta$,
at the pole we get back the expression (\ref{eq:31}). Then written
as in  (\ref{R}), $R$ and analogously $\Delta_{eff}$ become
functions of the velocity and the energy; therefore our average
should be better taken in the momentum space.

\section{Smearing for generic crystalline structures}\label{sec:GCS}
We have defined in the previous section a smearing procedure which,
in the case of the FF state, allows to write in the HDET formalism a
gap equation which coincides with the exact gap equation in
Eq.~\eqref{GAPeff}. Apart the technical details for the definition
of the weighting function, the smearing procedure can be view as a
recipe which allows to replace an ${\bf r}$-dependent gap function
by an ${\bf r}$-independent one. All the details of the crystalline
structure are embodied into the definition of the pairing regions.

We now use this recipe to smear the interaction lagrangian for
generic crystalline structures, defined by the pairing ansatz \be
\Delta(\bm r)=\Delta\sum_{m=1}^P e^{2i{\bf q_m}\cdot{\bf
r}}~,~~~~~{\bf q_m}\,
 =\,q\,{\bf n_m}~.
\label{de}\ee This procedure, although artificial, allows to study
generic crystalline structures in a domain where Ginzburg-Landau
expansion may not work well. Generalizing the results of the
previous equations we substitute in the Lagrangian (\ref{loff60bis})
$\Delta_{eff}\left({\bf v\cdot {\bf n}},\ell_0\right)$ with
\be\Delta_{E}({\bf v},\ell_0)=\sum_{m=1}^P\Delta_{eff}\left({\bf
v\cdot\bf n_m},\ell_0\right)\label{eq13}   ~ , \ee and the arguments
of the $\theta$ functions in Eq.~\eqref{GAP8.1} will be the
appropriate fermion dispersion laws. After smearing the gap equation
reads \be P \Delta=i~\frac{g\rho}{2}\int\frac{d{\bf v}}{4\pi}\int
\frac{d^2\ell}{2\pi} \, \frac{\Delta_{E}({\bf
v},\ell_0)}{\ell_0^2-\ell_\parallel^2-\Delta_{E}^2({\bf
v},\ell_0)}\label{29} \ \ee which generalizes Eq. (\ref{GAPeff}).
The energy integration is performed by the residue theorem and
 the phase space is divided into different regions according to
 the pole positions. We get \be
 P
\Delta\ln\frac{2\delta}{\Delta_0} =\sum_{k=1}^P \int\int_{P_k}
\frac{d{\bf v}}{4\pi} d\xi \, \frac{\Delta_{E}({\bf
v},\epsilon)}{\sqrt{\xi^2+\Delta^2_{E}({\bf v},\epsilon)}}
 =\sum_{k=1}^P\int\int_{P_k}\frac{d{\bf v}}{4\pi} d\xi \,
\frac{k\Delta}{\sqrt{\xi^2+k^2\Delta^2}}\label{290} \ \ee where the
pairing regions $P_k$ are defined as follows \be P_k=\{({\bf
v},\xi)\,|\,\Delta_E({\bf v},\epsilon)=k\Delta\}\ \ee and we have
ruled out the coupling constant $G$ by means of the BCS gap
$\Delta_0$. The first term in the sum, corresponding to the region
$P_1$, has $P$ equal contributions with a dispersion rule equal to
the Fulde and Ferrel case. This can be interpreted as a contribution
from $P$ non interacting plane waves. In the other regions the
different plane waves have an overlap.

Thus for $N$ plane waves the smearing procedure does not simply
leads to a lagrangian which is the sum of $N$ plane waves (FF)
lagrangians. This ``would be nice'' scenario is complicated by the
presence of the pairing regions with two or more overlaps
$(P_2,\dots,P_N)$.

\section{Numerical results for the free energy}\label{sec:NR}
\begin{figure}[t] \centerline{
\epsfxsize=5.5cm\epsfbox{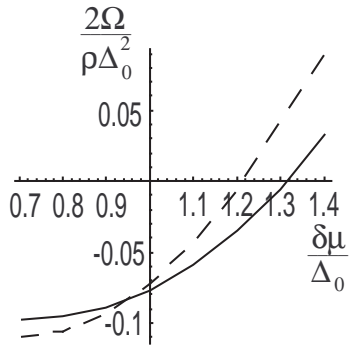} } \caption {{ \rm The values of
the free energies of the bcc (dashed line) and of the fcc (full
line) crystalline LOFF  structures as a function of
$\delta\mu/\Delta_0$. The bcc is the favored structure up to
$\delta\mu \approx .95 \Delta_0$; for $.95 \Delta_0< \delta\mu <
1.32 \Delta_0$ the fcc is favored. Here, for each value of
$\delta\mu$, the values of $z_q$ and $\Delta$ are those that
minimize the free energy. From Ref.~\cite{Casalbuoni:2004wm}.
\label{omegaplot} }}
\end{figure}

In this section we present results for the free energy $\Omega$ of
the structures that we have considered in
Ref.~\cite{Casalbuoni:2004wm}, computed as integral of the gap
equation~\eqref{290}. In particular, in Fig.~\ref{omegaplot} we show
$\Omega(\Delta,\delta\mu)-\Omega(0,\delta\mu)$  against $\delta\mu$
for the face centered (fcc) and the body centered (bcc) cubic
structures (the other structures have a higher free energy and
therefore they are not shown here). Our central result is that the
bcc is a good candidate for the ground state for $\delta\mu\le
0.95\Delta_0$. Above this value of $\delta\mu$ the fcc is the good
ground state for $\delta\mu\le 1.32\Delta_0$. For higher values of
$\delta\mu$, the fermion condensation is energetically disfavored.
The transition to the normal state is first order.

\section{Conclusions}\label{sec:CO}
In the framework of HDET, we have discussed an averaging procedure
of the NJL quark-quark interaction lagrangian, treated in the mean
field approximation, for the LOFF phase of QCD. This procedure gives
results which are valid in domains where Ginzburg-Landau results may
be questionable. Among the several structures considered, we find
that a body centered cube is the favorite ground state for
$\delta\mu\le0.95\Delta_0$, while for higher values of $\delta\mu$
the face centered cube is favoured. A first order transition to the
normal state is found for $\delta\mu\approx1.32\Delta_0$. The method
exposed here can be applied also to the LOFF phase in the
electromagnetic superconductors (see for example
Ref.~\cite{Casalbuoni:2004zy}).

\vspace{1cm}{\em Acknowledgments.}
  I am in debt with R.~Casalbuoni, R.~Gatto, N.~Ippolito, G.~Nardulli and M.~Ciminale for
  fruitful collaboration; moreover I thank M.~Alford, A.~Gerhold, M.~Mannarelli and
  I.~Shovkovy for clarifying discussions, comments and helpful correspondence.


\begin{thebibliography}{99}

%\cite{Collins:1974ky}
\bibitem{Collins:1974ky}
  J.~C.~Collins and M.~J.~Perry,
  %``Superdense Matter: Neutrons Or Asymptotically Free Quarks?,''
  Phys.\ Rev.\ Lett.\  {\bf 34}, 1353 (1975).
  %%CITATION = PRLTA,34,1353;%%

%\cite{Barrois:1977xd}
\bibitem{Barrois:1977xd}
  B.~C.~Barrois,
  %``Superconducting Quark Matter,''
  Nucl.\ Phys.\ B {\bf 129}, 390 (1977).
  %%CITATION = NUPHA,B129,390;%%

  %\cite{Bailin:1983bm}
\bibitem{Bailin:1983bm}
  D.~Bailin and A.~Love,
  %``Superfluidity And Superconductivity In Relativistic Fermion Systems,''
  Phys.\ Rept.\  {\bf 107}, 325 (1984).
  %%CITATION = PRPLC,107,325;%%

%\cite{Alford:1997zt}
\bibitem{Alford:1997zt}
  M.~G.~Alford, K.~Rajagopal and F.~Wilczek,
  %``QCD at finite baryon density: Nucleon droplets and color
  %superconductivity,''
  Phys.\ Lett.\ B {\bf 422}, 247 (1998)
  [arXiv:hep-ph/9711395].
  %%CITATION = HEP-PH 9711395;%%


%\cite{Rapp:1997zu}
\bibitem{Rapp:1997zu}
  R.~Rapp, T.~Schafer, E.~V.~Shuryak and M.~Velkovsky,
  %``Diquark Bose condensates in high density matter and instantons,''
  Phys.\ Rev.\ Lett.\  {\bf 81}, 53 (1998)
  [arXiv:hep-ph/9711396].
  %%CITATION = HEP-PH 9711396;%%

%\cite{Schafer:1999jg}
\bibitem{Schafer:1999jg}
  T.~Schafer and F.~Wilczek,
  %``Superconductivity from perturbative one-gluon exchange in high density
  %quark matter,''
  Phys.\ Rev.\ D {\bf 60}, 114033 (1999)
  [arXiv:hep-ph/9906512].
  %%CITATION = HEP-PH 9906512;%%



  %\cite{Alford:1998mk}
\bibitem{Alford:1998mk}
  M.~G.~Alford, K.~Rajagopal and F.~Wilczek,
  %``Color-flavor locking and chiral symmetry breaking in high density {QCD},''
  Nucl.\ Phys.\ B {\bf 537}, 443 (1999)
  [arXiv:hep-ph/9804403].
  %%CITATION = HEP-PH 9804403;%%

  %\cite{Shovkovy:2003uu}
\bibitem{Shovkovy:2003uu}
  I.~Shovkovy and M.~Huang,
  %``Gapless two-flavor color superconductor,''
  Phys.\ Lett.\ B {\bf 564}, 205 (2003)
  [arXiv:hep-ph/0302142].
  %%CITATION = HEP-PH 0302142;%%

%\cite{Alford:2003fq}
\bibitem{Alford:2003fq}
  M.~Alford, C.~Kouvaris and K.~Rajagopal,
  %``Gapless color-flavor-locked quark matter,''
  Phys.\ Rev.\ Lett.\  {\bf 92}, 222001 (2004)
  [arXiv:hep-ph/0311286].
  %%CITATION = HEP-PH 0311286;%%

%\cite{Rajagopal:2000wf}
\bibitem{Rajagopal:2000wf}
  K.~Rajagopal and F.~Wilczek,
  %``The condensed matter physics of QCD,''
  arXiv:hep-ph/0011333.
  %%CITATION = HEP-PH 0011333;%%

%\cite{Alford:2001dt}
\bibitem{Alford:2001dt}
  M.~G.~Alford,
  %``Color superconducting quark matter,''
  Ann.\ Rev.\ Nucl.\ Part.\ Sci.\  {\bf 51}, 131 (2001)
  [arXiv:hep-ph/0102047].
  %%CITATION = HEP-PH 0102047;%%

  %\cite{Nardulli:2002ma}
\bibitem{Nardulli:2002ma}
  G.~Nardulli,
  %``Effective description of QCD at very high densities,''
  Riv.\ Nuovo Cim.\  {\bf 25N3}, 1 (2002)
  [arXiv:hep-ph/0202037].
  %%CITATION = HEP-PH 0202037;%%

%\cite{Schafer:2003vz}
\bibitem{Schafer:2003vz}
  T.~Schafer,
  %``Quark matter,''
  arXiv:hep-ph/0304281.
  %%CITATION = HEP-PH 0304281;%%

%\cite{Schaefer:2005ff}
\bibitem{Schaefer:2005ff}
  T.~Schaefer,
  %``Phases of QCD,''
  arXiv:hep-ph/0509068.
  %%CITATION = HEP-PH 0509068;%%

\bibitem{LOFF1}
A. I. Larkin,  Yu. N. Ovchinnikov, Zh. Eksp. Teor. Fiz.47, (1964),
1136.

\bibitem{LOFF2}
P. Fulde, R. A. Ferrel, Phys. Rev.135 (1964) ,A550.


%\cite{Alford:2000ze}
\bibitem{Alford:2000ze}
  M.~G.~Alford, J.~A.~Bowers and K.~Rajagopal,
  %``Crystalline color superconductivity,''
  Phys.\ Rev.\ D {\bf 63}, 074016 (2001)
  [arXiv:hep-ph/0008208].
  %%CITATION = HEP-PH 0008208;%%

%\cite{Bowers:2001ip}
\bibitem{Bowers:2001ip}
  J.~A.~Bowers, J.~Kundu, K.~Rajagopal and E.~Shuster,
  %``A diagrammatic approach to crystalline color superconductivity,''
  Phys.\ Rev.\ D {\bf 64}, 014024 (2001)
  [arXiv:hep-ph/0101067].
  %%CITATION = HEP-PH 0101067;%%

%\cite{Bowers:2002xr}
\bibitem{Bowers:2002xr}
  J.~A.~Bowers and K.~Rajagopal,
  %``The crystallography of color superconductivity,''
  Phys.\ Rev.\ D {\bf 66}, 065002 (2002)
  [arXiv:hep-ph/0204079].
  %%CITATION = HEP-PH 0204079;%%

%\cite{Casalbuoni:2003wh}
\bibitem{Casalbuoni:2003wh}
  R.~Casalbuoni and G.~Nardulli,
  %``Inhomogeneous superconductivity in condensed matter and QCD,''
  Rev.\ Mod.\ Phys.\  {\bf 76}, 263 (2004)
  [arXiv:hep-ph/0305069].
  %%CITATION = HEP-PH 0305069;%%

%\cite{Giannakis:2004pf}
\bibitem{Giannakis:2004pf}
  I.~Giannakis and H.~C.~Ren,
  %``Chromomagnetic instability and the LOFF state in a two flavor color
  %superconductor,''
  Phys.\ Lett.\ B {\bf 611}, 137 (2005)
  [arXiv:hep-ph/0412015].
  %%CITATION = HEP-PH 0412015;%%

%\cite{Giannakis:2005sa}
\bibitem{Giannakis:2005sa}
  I.~Giannakis, D.~f.~Hou and H.~C.~Ren,
  %``A neutral two flavor LOFF color superconductor,''
  arXiv:hep-ph/0507306.
  %%CITATION = HEP-PH 0507306;%%

  %\cite{Giannakis:2005vw}
\bibitem{Giannakis:2005vw}
  I.~Giannakis and H.~C.~Ren,
  %``The Meissner effect in a two flavor LOFF color superconductor,''
  Nucl.\ Phys.\ B {\bf 723}, 255 (2005)
  [arXiv:hep-th/0504053].
  %%CITATION = HEP-TH 0504053;%%

%\cite{Casalbuoni:2005zp}
\bibitem{Casalbuoni:2005zp}
  R.~Casalbuoni, R.~Gatto, N.~Ippolito, G.~Nardulli and M.~Ruggieri,
  %``Ginzburg-Landau approach to the three flavor LOFF phase of QCD,''
  arXiv:hep-ph/0507247.
  %%CITATION = HEP-PH 0507247;%%

  %\cite{Hong:1998tn}
\bibitem{Hong:1998tn}
  D.~K.~Hong,
  %``An effective field theory of {QCD} at high density,''
  Phys.\ Lett.\ B {\bf 473}, 118 (2000)
  [arXiv:hep-ph/9812510].
  %%CITATION = HEP-PH 9812510;%%

%\cite{Hong:1999ru}
\bibitem{Hong:1999ru}
  D.~K.~Hong,
  %``Aspects of high density effective theory in {QCD},''
  Nucl.\ Phys.\ B {\bf 582}, 451 (2000)
  [arXiv:hep-ph/9905523].
  %%CITATION = HEP-PH 9905523;%%

%\cite{Beane:2000ms}
\bibitem{Beane:2000ms}
  S.~R.~Beane, P.~F.~Bedaque and M.~J.~Savage,
  %``Meson masses in high density QCD,''
  Phys.\ Lett.\ B {\bf 483}, 131 (2000)
  [arXiv:hep-ph/0002209].
  %%CITATION = HEP-PH 0002209;%%

%\cite{Schafer:2003jn}
\bibitem{Schafer:2003jn}
  T.~Schafer,
  %``Hard loops, soft loops, and high density effective field theory,''
  Nucl.\ Phys.\ A {\bf 728}, 251 (2003)
  [arXiv:hep-ph/0307074].
  %%CITATION = HEP-PH 0307074;%%

%\cite{Casalbuoni:2001dw}
\bibitem{Casalbuoni:2001dw}
  R.~Casalbuoni,
  %``Effective Lagrangians for QCD at high density,''
  AIP Conf.\ Proc.\  {\bf 602}, 358 (2001)
  [arXiv:hep-th/0108195].
  %%CITATION = HEP-TH 0108195;%%

  %\cite{Schafer:2003yh}
\bibitem{Schafer:2003yh}
  T.~Schafer,
  %``Loops and power counting in the high density effective field theory,''
  eConf {\bf C030614}, 038 (2003)
  [arXiv:hep-ph/0310176].
  %%CITATION = HEP-PH 0310176;%%

  %\cite{Casalbuoni:2004wm}
\bibitem{Casalbuoni:2004wm}
  R.~Casalbuoni, M.~Ciminale, M.~Mannarelli, G.~Nardulli, M.~Ruggieri and R.~Gatto,
  %``Effective gap equation for the inhomogeneous LOFF superconductive  phase,''
  Phys.\ Rev.\ D {\bf 70}, 054004 (2004)
  [arXiv:hep-ph/0404090].
  %%CITATION = HEP-PH 0404090;%%

  %\cite{Leibovich:2001xr}
\bibitem{Leibovich:2001xr}
  A.~K.~Leibovich, K.~Rajagopal and E.~Shuster,
  %``Opening the crystalline color superconductivity window,''
  Phys.\ Rev.\ D {\bf 64}, 094005 (2001)
  [arXiv:hep-ph/0104073].
  %%CITATION = HEP-PH 0104073;%%

  %\cite{Casalbuoni:2003sa}
\bibitem{Casalbuoni:2003sa}
  R.~Casalbuoni, R.~Gatto, M.~Mannarelli, G.~Nardulli, M.~Ruggieri and S.~Stramaglia,
  %``Quasi-particle specific heats for the crystalline color superconducting
  %phase of QCD,''
  Phys.\ Lett.\ B {\bf 575}, 181 (2003)
  [Erratum-ibid.\ B {\bf 582}, 279 (2004)]
  [arXiv:hep-ph/0307335].
  %%CITATION = HEP-PH 0307335;%%

%\cite{Casalbuoni:2004zy}
\bibitem{Casalbuoni:2004zy}
  R.~Casalbuoni, R.~Gatto, M.~Mannarelli, G.~Nardulli and M.~Ruggieri,
  %``Magnetic properties of the Larkin-Ovchinnikov-Fulde-Ferrell
  %superconducting phase,''
  Phys.\ Lett.\ B {\bf 600}, 48 (2004)
  [arXiv:hep-ph/0407210].
  %%CITATION = HEP-PH 0407210;%%




\end{thebibliography}
\end{document}